\documentstyle[prl,aps,epsf]{revtex}
\newcommand{\be}{\begin{equation}}
\newcommand{\ee}{\end{equation}}
\newcommand\mathnew{\mathsurround=0pt}
\def\simov#1#2{\lower .5pt\vbox{\baselineskip0pt \lineskip-.5pt
        \ialign{$\mathnew#1\hfil##\hfil$\crcr#2\crcr\sim\crcr}}}
\newcommand\simg{\mathrel{\mathpalette\simov >}}
\newcommand\siml{\mathrel{\mathpalette\simov <}}

\begin{document}

\preprint{LAUR 97-XXXX, CGPG-97/X-X}

\draft
\twocolumn

\title{Critical Dynamics of Symmetry Breaking: \\
       Quenches, Dissipation and Cosmology}
\author{Pablo Laguna${}^{(1,2)}$ and Wojciech Hubert Zurek${}^{(1)}$}
\address{
${}^{(1)}$ Theoretical Astrophysics, MS-B288,
Los Alamos Nat. Lab., Los Alamos, NM 87545, USA\\
${}^{(2)}$ Dept. Astronomy \& Astrophysics and CGPG,
Penn State University, University Park, PA 16802, USA}
\date{\today}
\maketitle
\begin{abstract}
Symmetry-breaking phase transitions may leave behind topological 
defects \cite{Kibble} with a density dependent on the quench rate \cite{Zurek}.
We investigate the dynamics of such quenches for the one-dimensional,
Landau-Ginzburg case and show that the density of kinks, $n$, scales
differently with the quench timescale, $\tau_Q$, depending on
whether the dynamics in the vicinity of the critical point is
overdamped ($n \propto \tau_Q^{-1/4}$) or underdamped
($n \propto \tau_Q^{-1/3}$).  Either of these cases may be relevant to the early
Universe, and we derive bounds on the initial density of topological
defects in cosmological phase transitions.
\end{abstract}
\pacs{05.70.Fh, 11.15.Ex, 67.40.Vs, 74.60-w}

Dynamics of symmetry breaking phase transformations is of interest in 
cosmology, condensed matter and high energy physics.
Recent surge of interdisciplinary interest has been fueled by the
experiments involving creation of topological defects during
rapid second order phase transitions in superfluids 
\cite{Hendry,Ruutu,Bauerle,Helsinki}
and other systems \cite{Chuang}, in a setting reminiscent of similar
cosmological processes \cite{Kibble,Zeldovich}. 

In the cosmological context,
topological defects such as cosmic strings may have played a role in seeding
structure formation \cite{Kibble,Vilenkin}.
In high energy physics, accelerator experiments may allow one to probe
restoration of some of the symmetries (e.g. chiral), which were originally
broken early in the history of the Universe.
The signature of whether such restoration has occurred will come 
from the fluctuation of decay products \cite{Bjorken,Rajagopal}, determined
by the relevant dynamics of the order parameter during a quench.
In superfluids, the very same process controls the production of topological 
defects. The interest in  experimental exploration of the critical dynamics is
therefore well-justified by its wide-ranging applications, which
may also come to include in the near future the creation of
vortices in Bose-Einstein condensates \cite{BEC}.

These processes span many orders of magnitude in spatial and energetic scales.  
Yet, a large class of them is well-approximated by the Landau-Ginzburg 
theory. Therefore, the dynamics of the order parameter $\varphi$
is governed by an equation of the form \cite{note1}:
\be
\ddot \varphi + \eta\, \dot \varphi -
c^2 \nabla^2 \varphi + [\beta\,\varphi^3 
- m^2\,\epsilon(t)\,\varphi]/2 = \vartheta(t,x) \, .
\label{eq:1}
\ee
Above, $\eta$ characterizes viscosity, while $c,\, \beta$
and $m$ are constant coefficients, and
$\epsilon(t)$ is the time-dependent relative temperature,
assumed to vary with time as
$
\epsilon = t/\tau_Q \, ,
$
where $\tau_Q$ the is quench timescale. 
The term $\vartheta(t,x)$ is noise characterized 
by its spatial and temporal correlations, 
as well as by its amplitude $\theta$.
We assume:
$
\langle \vartheta (x, t), \vartheta (x', t') \rangle =  
2\, \eta\, \theta\, \delta(x'-x)
\,\delta(t'-t)$. 
Equation~(\ref{eq:1}) can be expressed in
``natural" units $t \rightarrow t/m,\, 
x \rightarrow x\, c/m,\, \eta \rightarrow \eta\, m, \,
\varphi \rightarrow \varphi\, m/\sqrt{\beta}$ and
$\theta \rightarrow \theta\, m^3\, c/ \beta$,
which leads to:
\be
\ddot \varphi + \eta\, \dot \varphi -
\nabla^2 \varphi + (\varphi^3-\epsilon\varphi)/2 = \vartheta \, .
\label{eq:3}
\ee
Thus, in the vicinity of the second-order phase transition, 
an enormous range of ``bare" parameters can be reduced to 
two: the ``renormalized" damping rate $\eta$ and the noise 
temperature $\theta$. The quench adds a dependence of the
consequences of critical dynamics on 
the quench rate $\dot \epsilon = \tau_Q^{-1}$.
The aim of our study is to investigate the dependence of the
critical dynamics on the value of $\eta$ and $\tau_Q$, 
under the assumption that $\theta$ is sufficiently small,
so the probability of thermally activated symmetry restoration
is negligible after the quench. In this paper we take $\theta = 0.01$.
We focus on creation of kinks --- one-dimensional
topological defects --- in the course of rapid quenches.

Evolution generated by Eq.~(\ref{eq:3}) is overdamped when
$\eta\,\dot \varphi > \ddot \varphi$. In this regime, 
the relaxation time $\tau_{\dot \varphi} \simeq |\varphi/\dot\varphi|$ 
scales with the relative temperature
$\epsilon$ as:
$
\tau_{\dot \varphi} \simeq \eta\,\tau_o^2\,|\epsilon|^{-1} 
\simeq \eta\,\tau_Q\,\tau_o^2\,|t|^{-1}\, ,
$
in the units of Eq.~(\ref{eq:1}) 
with $\tau_o = m^{-1}$ the dynamical timescale.
In accord with \cite{Zurek}, one expects the initial size
$\hat\xi$ of the pieces of the new broken symmetry phase to be set at the 
time $\hat t$, when the time to (from) the phase 
transition is comparable to the relaxation timescale \cite{Zurek},
and the freeze-out of the field evolution occurs; that is, its state
cannot keep up with the change of the thermodynamic parameters as a result of
critical slowing down. This 
freeze-out condition,
$\tau_{\dot \varphi}(\hat t_{\dot \varphi}) = \hat t_{\dot \varphi}$
yields
\begin{eqnarray}
\label{eq:5}
\hat t_{\dot \varphi} & \simeq & \tau_o\,(\eta\,\tau_Q)^{1/2} \\
\label{eq:6} 
\hat\epsilon_{\dot \varphi} & \simeq  &
\left(\frac{\eta\,\tau_o^2}{\tau_Q}\right)^{1/2} \, .
\end{eqnarray}
The correlation length $\hat\xi$, which sets the stage for
the defect formation \cite{Kibble}, is then:
\be
\hat\xi_{\dot \varphi} \simeq 
\frac{\xi_o}{|\hat\epsilon_{\dot \varphi}|^{1/2}}
\simeq \xi_o \left(\frac{\tau_Q}{\eta\,\tau_o^2}\right)^{1/4} \, ,
\label{eq:7}
\ee
where $\xi_o = c/m$ characterizes the low temperature
($\epsilon = 0$) healing length.

To test these arguments \cite{Zurek}, we have recently carried out a
numerical study of defect formation \cite{Laguna}, showing that
the density of kinks formed in a quench indeed varies,
in the overdamped regime, as:
\be
n_{\dot \varphi} \simeq \frac{1}{f\hat\xi_{\dot \varphi}}
\propto \left(\frac{\eta\,\tau_o^2}{\tau_Q}\right)^{1/4}
\label{eq:8}
\ee
with $f \simeq 8$.
Similar study was independently carried out by Lythe \cite{Lythe},
who has estimated $f = 2\pi(\ln{\theta})^{1/4}$
when $\theta \ll 1$.

Our purpose here is to extend these studies from the regime 
where damping dominates (which is most relevant in condensed matter
applications) to the range where the evolution is
underdamped (as may be the case in cosmology).
For details of the numerical technique, see
Ref.~\cite{Laguna}. 

In the underdamped case,  
$\ddot \varphi$ will dominate, and the order parameter
reacts to the quench-induced changes in the 
effective potential on the timescale
$\tau_{\ddot \varphi} \simeq |\varphi/\ddot\varphi|^{1/2}$. Thus,
$
\tau_{\ddot \varphi} \simeq
\tau_o\,|\epsilon|^{-1/2} \, .
$
The freeze-out condition, 
$\tau_{\ddot \varphi}(\hat t_{\ddot \varphi}) = \hat t_{\ddot \varphi}$ 
yields in this underdamped regime:
\begin{eqnarray}
\hat t_{\ddot \varphi} & \simeq & \tau_o (\tau_Q/\tau_o)^{1/3} \, ;\\
\hat \epsilon_{\ddot \varphi} & \simeq & (\tau_o/\tau_Q)^{2/3} \, .
\label{eq:5b}
\end{eqnarray}
Consequently, the scaling of the characteristic correlation length
with the quench rate $\tau_Q^{-1}$ is expected to change to
\be
\hat\xi_{\ddot \varphi} \simeq 
\frac{\xi_o}{|\hat\epsilon_{\ddot \varphi}|^{1/2}}
\simeq \xi_o \left(\frac{\tau_Q}{\tau_o}\right)^{1/3} \, .
\label{eq:7b}
\ee
Furthermore, the density of the number of kinks is given in this case by:
\be
n_{\ddot \varphi} \simeq \frac{1}{f\hat\xi_{\ddot \varphi}}
\propto \left(\frac{\tau_o}{\tau_Q}\right)^{1/3} \, ,
\label{eq:8b}
\ee
although $f$ may be now different.

We can therefore draw two related conclusions: (i)
In the overdamped regime, the density of kinks should scale with
$\eta^{1/4}$, and should become viscosity independent in the
underdamped case. (ii) Power-law dependence of the density of kinks
with the quench timescale should change from $\propto \tau_Q^{-1/4}$
in the overdamped case to $\propto \tau_Q^{-1/3}$ in the
underdamped case. The overdamped scalings should apply
when the evolution is dominated by the first derivative 
($\eta\,\dot \varphi > \ddot \varphi$, i.e. 
$\eta/\tau_{\dot\varphi} > 1/\tau_{\ddot\varphi}^2$)
at the instant when topological defects ``freeze-out."
This will happen for:
$
|\hat\epsilon_{\dot\varphi}| > |\hat\epsilon_{\ddot\varphi}| \, ,
$
or --- using Eqs.~(\ref{eq:6}) and (\ref{eq:5b}) --- when:
\be
(\eta\,\tau_o)^{3} > (\tau_o/\tau_Q) \, .
\label{eq:10}
\ee

\begin{figure}[h]
\leavevmode
\epsfxsize=3.2truein\epsfbox{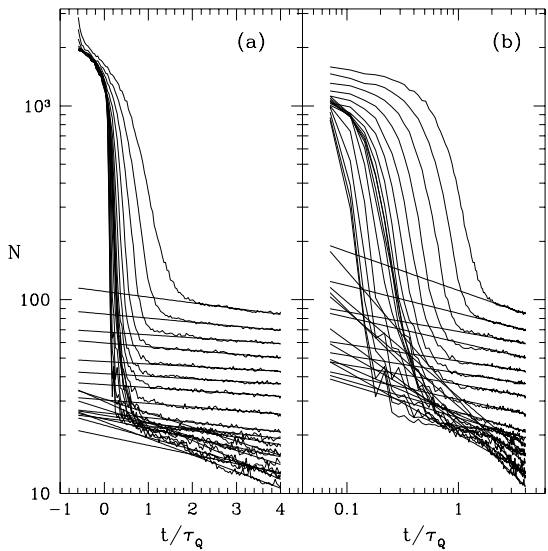}
\caption[figure1]{\label{fig1}
Number of kinks $N$ as a function of time for simulations with different 
viscosity parameters $\eta$, but with the same quench timescale $\tau_Q=256$.
Two models for the annihilation rate are shown, (a) exponential
$N = N_o \exp{(-a\,t/\tau_Q)}$ and (b) power-law $N = N_o (t/\tau_Q)^{-b}$.
Note the increase of the annihilation rates for small $\eta$.
}
\end{figure}

\begin{figure}[h]
\leavevmode
\epsfxsize=3.2truein\epsfbox{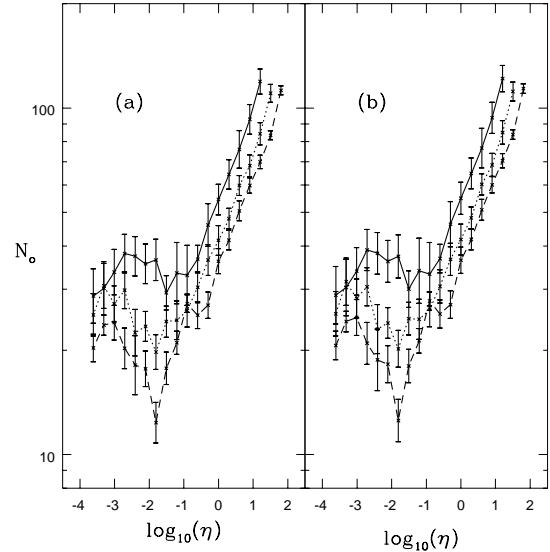}
\caption[figure2]{\label{fig2}
Initial number of kinks $N_o$ as a function of the 
damping rate $\eta$ for a fixed quench rate
timescales (top to bottom $\tau_Q =$ 128, 256 and 512).
Both exponential (a) and power-law (b) model results (see Fig.~\ref{fig1}) are shown
(and are essentially identical).
For $\eta > 0.1$, $N_o \propto \eta^g$, where
$N_o$ is obtained from the fittings in Fig~\ref{fig1}.
Best fits yield 
$g = (0.27 \pm 0.035,\,
0.25 \pm 0.029,\, 0.27 \pm 0.011)$ for
$\tau_Q = (128,\, 256,\, 512)$, respectively.
}
\end{figure}

We identify kinks as zeros of the order parameter. 
This can be justified only well after the phase transition,
when $\varphi$ has locally settled into the broken symmetry state.
Kinks annihilate, and their number slowly decreases with time.
Previously, in the overdamped regime,
we were able to confirm the predicted \cite{Zurek}
dependence of the initial number of
kinks on $\tau_Q$ from the numerical data by using a 
fairly straightforward procedure of simply counting zeros at a fixed value of 
$t/\tau_Q$ \cite{Laguna}. The nature of that dependence did not change
dramatically even when the counting of kinks was taking
at a constant value of $t$ (although a change on the slope
as well as evidence of the saturation in the number of 
kinks for small $\tau_Q$ were noted).
But the nature of critical dynamics and especially the 
annihilation rate depend on $\eta$, 
which we shall vary by several orders of magnitude.
To compare ``initial densities" of kinks now, we therefore
need a more objective procedures independent of the time
at which the kinks are counted. We have done this by using
whole runs of kink densities (such as the ones shown in Fig.~\ref{fig1})
to model annihilation either as a power-law
$N \simeq N_o  (t/\tau_Q)^{-b}$, or as 
an exponential, $N \simeq N_o \exp{(-a\, t/\tau_Q)}$, with $N$
the number of zeros of the order parameter.
The actual dependences are usually sufficiently similar to a
straight line that both of these procedures yield reasonable
initial kink numbers $N_o$, and at least in some cases an almost surprisingly
similar dependences on the parameters (see Fig.~\ref{fig2}).

The dependence of the initial number of kinks $N_o$ 
on the damping coefficient $\eta$ for three different
quench rates ($\tau_Q = 128,\, 256,\, \hbox{and}\, 512$)
is shown in Fig.~\ref{fig2}.
In the regime of large viscosities,
critical dynamics is overdamped, in accord with 
Eq.~(\ref{eq:10}), and leads to the power-law 
dependence $\propto \eta^{1/4}$, Eq.~(\ref{eq:8}).
The spacing between the three lines is also roughly
consistent with the one anticipated from that equation.
As the damping rate decreases below the value estimated from 
Eq.~(\ref{eq:10}), namely $\eta \siml 0.1$,
the number of kinks becomes essentially 
independent of $\eta$.
Moreover, the spacing between the (now approximately
horizontal) lines of constant $\tau_Q$ is consistent with
the underdamped case $N_o \propto \tau_Q^{-1/3}$, Eq.~(\ref{eq:8b}).

We should note, however, that for $\eta \le 10^{-2}$, 
the number of kinks are small and the annihilation is more efficient.
Consequently, our results in this range are less reliable. 
In a sense, Fig.~\ref{fig1} indicates a more systematic (and probably more consistent 
with the theoretical expectations) trend with the damping rate $\eta$
than our estimates of the number of kinks plotted in Fig.~\ref{fig2}. Most of the 
scatter in Fig.~\ref{fig2} stems from our inability to model annihilation rate in 
a consistent fashion over a broad range of parameters, rather than from
the ``raw'' data shown in Fig.~\ref{fig1}.

\begin{figure}[h]
\leavevmode
\epsfxsize=3.2truein\epsfbox{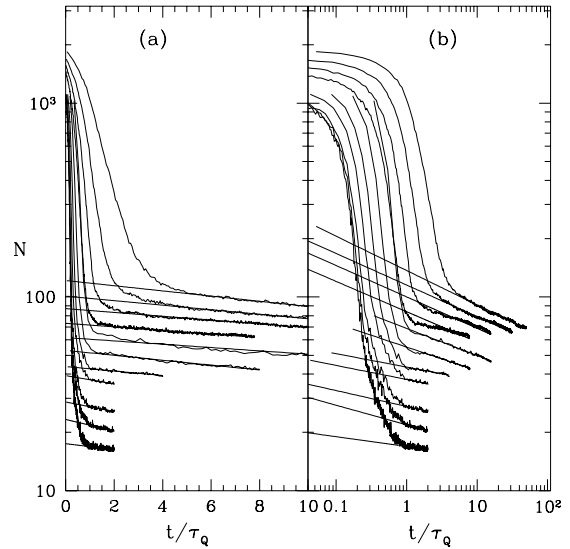}
\caption[figure3]{\label{fig3}
Number of kinks $N$ as a function of time for a fixed viscosity
$\eta = 1$ but different quench timescales $\tau_Q$.
As with Fig~\ref{fig1},
(a) corresponds to an exponential fit and (b) to a power-law fit.
Saturation is apparent in the case (a) for short quench timescales (and large initial
kink densities).
}
\end{figure}

\begin{figure}[h]
\leavevmode
\epsfxsize=3.2truein\epsfbox{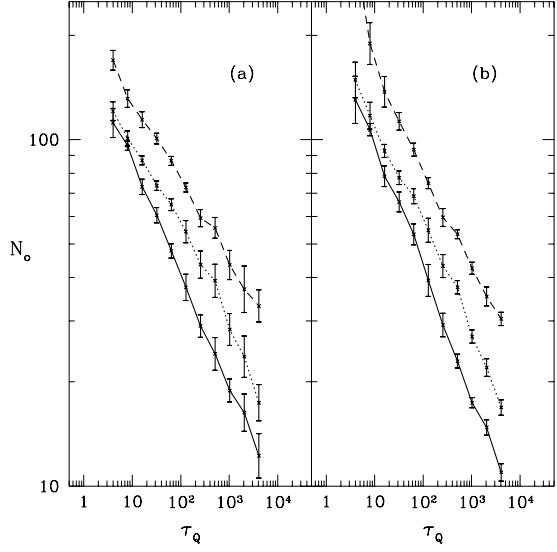}
\caption[figure4]{\label{fig4}
Dependence of the initial number
of kinks $N_o$ on the quench timescale $\tau_Q$ for 
values of the damping rate 
$\eta$ = 5 (top), 1 (middle) and 1/5 (bottom).
Case (a) is obtained from an exponential fitting to the
decay of number of kinks and (b) from a power-law fit.
Fittings to $N_o \propto \tau_Q^{-g}$ yield (from top to
bottom) $g = (0.23 \pm 0.010,\,
0.26 \pm 0.011,\, 0.33 \pm 0.011)$ in case (a)
and $g = (0.28 \pm 0.010,\,
0.30 \pm 0.011,\, 0.36 \pm 0.010)$ in case (b).
}
\end{figure}

These conclusions concerning the transition from overdamped to
underdamped behavior are strengthened by comparing families
of simulations corresponding to the same $\eta$
but for varying $\tau_Q$ (see Fig.~\ref{fig3} and \ref{fig4}).
As before, we consider two methods for obtaining the initial 
number of kinks $N_o$ from the
time-dependent data, Fig.~\ref{fig3}. 
In the range of long quench timescales they produce
similar (but not identical) power-laws.

According to condition~(\ref{eq:10}), the $\eta = 5$ and 1 cases in Fig.~\ref{fig4} 
are, for the values of $\tau_Q$ under consideration ($2 \le \tau_Q \le 4098$), entirely 
within the overdamped regime.
For these two cases, we find power-laws $\sim \tau_Q^{-1/4}$,
consistent with Eq.~(\ref{eq:8}). On the other hand, for the $\eta = 1/5$ case
in accord with Eq.~(\ref{eq:10}), 
a transition between overdamped and underdamped regimes should occur at
$\tau_Q \sim 125$. We find 
(see Fig.~\ref{fig4}) indication of a change in the power-law dependence,
from $N_o \propto \tau_Q^{-1/4}$ to $\propto \tau_Q^{-1/3}$ as $\eta$
decreases. 

One obvious case of breakdown of power-laws, 
Eqs.~(\ref{eq:8}) and (\ref{eq:8b}), occurs when the condition
$\hat\epsilon \ll  1$ of applicability of the theory of 
Ref.~\cite{Zurek} is not satisfied~\cite{ZZ}. In that case, the predicted
initial separation of kinks would be comparable to
(or even smaller than) the zero-temperature healing length $\xi_o$.
This would of course result in a rapid initial annihilation,
so that the density of defects would be set by the annihilation process
rather than by the critical dynamics.
We have seen evidence for this behavior for sufficiently
small $\tau_Q$ in Ref.~\cite{Laguna}.

Equations~(\ref{eq:8}) and (\ref{eq:8b}) for the initial
density of topological defects can be used in the
cosmological setting.
Phase transitions are likely to occur in the radiation
dominated era, when the temperature $T$ of the plasma and the 
Hubble time $t_H$ since the Big Bang are tied with the
equation $T^2\,t_H =$ constant. This immediately yields
quench timescales
$
\tau_Q = 2\,t_H = H^{-1}\, ,
$
where $H$ is the Hubble parameter.

Damping rate is the other important
parameter set by cosmology. In the radiation-dominated epoch
$
\eta = 3\,H + \gamma \, ,
$
where $3\,H$ is the effective viscosity caused by
the Hubble expansion, while $\gamma$ is damping due
to the coupling with the other degrees of freedom.
Early on, the ``Hubble viscosity" may even
dominate.

The nature of the critical dynamics in the immediate vicinity
of the phase transition is decided by the inequality~(\ref{eq:10}),
which now reads
\be
\left(\frac{3\,H+\gamma}{m}\right)^3 > \frac{H}{m} \, .
\label{eq:22}
\ee
In the overdamped case
\begin{eqnarray}
\hat t & \simeq & \tau_o\left(3 + \gamma/H\right)^{1/2} \\
\hat\epsilon & \simeq &\tau_o\,H\left(3+\gamma/H\right)^{1/2} \, .
\end{eqnarray}
This in turn leads to
\be
\hat\xi \simeq (\xi_o\, H^{-1})^{1/2}(3+\gamma/H)^{-1/4} \, , 
\label{eq:25}
\ee
where we have set $c = 1$, so  $\xi_o = m^{-1}$.
Thus the density of topological defects is principally set by the geometric average of 
the characteristic length scale of the 
order parameter ($\xi_o$)
and the size of the horizon ($H^{-1}$).
For small $\gamma/H$, damping is dominated by Hubble expansion.
In that regime, $H/m > 1$ (or $H^{-1} < \xi_o$) would be
required for the critical dynamics to be overdamped.
This would lead to $\hat\xi \simg H^{-1}$,
which in effect implies that in this case
the density of defects is set by the size of the horizon.

When the critical dynamics is underdamped;
\begin{eqnarray}
\hat t & \simeq & \tau_o\, (\tau_o\,H)^{-1/3} \, ,\\
\hat\epsilon & \simeq & (\tau_o\, H)^{2/3} \, .
\label{eq:27}
\end{eqnarray}
This immediately leads to
\be
\hat\xi \simeq \xi_o\,(H^{-1}/\xi_o)^{1/3}   
\label{eq:28}
\ee
The characteristic distance $\hat\xi$ is then
bigger than the healing length $\xi_o$
by the third root of the size of the horizon at the time
of the transition measured in the units of $\xi_o$.

We conclude by noting that the dependence of the number of kinks on the 
viscosity parameter corroborates anticipated existence of the two regimes
in the critical dynamics, each with a distinct scaling of the relevant
characteristic timescale 
with the relative temperature $\epsilon$. Overdamped regime produces
kinks with separations $\hat\xi \propto (\tau_Q/\eta)^{1/4}$, while in 
the underdamped case $\hat\xi \propto \tau_Q^{1/3}$ and is independent of 
viscosity. The subsequent annihilation rate of the kinks strongly depends on
viscosity, and is much more rapid in the underdamped case. The borderline 
between the two is consistent with the considerations of \cite{Zurek,ZZ}.

Work supported in part by 
NSF PHY 96-01413, 93-57219 (NYI) to PL.

%
%        figure 1
%

%
%        figure 2 
%

%
%        Figure 3
%
%
%        figure 4 
%

\end{document}